\documentclass[review]{elsarticle}

\usepackage{hyperref}

\journal{Nuclear Instruments and Methods in Physics A}

\biboptions{numbers,sort&compress}

\usepackage{subcaption}
\usepackage{cleveref}
\usepackage{booktabs}
\usepackage{multirow}

\begin{document}
\begin{frontmatter}

\title{High-precision characterization of the neutron light output of stilbene along the directions of maximum and minimum response}

\cortext[correspondingauthor]{Corresponding author}
\address[address]{Department of Nuclear Engineering, North Carolina State University, Raleigh, NC, USA}
\address[inrad_address]{Inrad Optics, 181 Legrand Avenue, Northvale, NJ , USA}
\address[duke]{Department of Physics, Duke University, Durham, NC, USA}
\address[patricia]{Department of Nuclear Engineering and Radiological Sciences, University of Michigan, Ann Arbor, MI, USA} 

\author[address]{R. A. Weldon Jr.\corref{correspondingauthor}}
\ead{raweldon@ncsu.edu}
\author[address]{J. M. Mueller}
\author[inrad_address]{C. Lynch}
\author[patricia]{P. Schuster}
\author[duke]{S. Hedges}
\author[duke]{C. Awe}
\author[duke]{L. Li}
\author[duke]{P. Barbeau}
\author[address]{J. Mattingly}

\begin{abstract}
The scintillation light output response of stilbene crystals has been measured for protons recoiling along the \textit{a}, \textit{b}, and \textit{c'} crystalline axes with energies between 1.3 and 10 MeV using neutrons produced with the tandem Van de Graaff accelerator at Triangle Universities Nuclear Laboratory.  The proton recoil energy and direction were measured using the coincident detection of neutrons between a stilbene scintillator and an array of EJ-309 liquid scintillators spanning arranged neutron recoil angles.  The maximum light output was found to coincide with proton recoils along the \textit{a}-axis, in disagreement with other published measurements, which reported the \textit{b}-axis as the direction of the maximum light output.  Additional measurements were conducted using two different stilbene crystals to confirm these results: a second measurement using the coincident detection of neutrons; measurements of neutron full energy deposition events along the \textit{a} and \textit{b} axes; and measurements of the count rate for $^{252}$Cf neutrons traveling along the \textit{a} and \textit{b} axes directions.  All measurements found that recoils along the \textit{a}-axis produce the maximum light output. 
\end{abstract}

\begin{keyword}
Stilbene\sep Scintillation anisotropy \sep Crystalline organic scintillator \sep Neutron detection \sep Detector characterization
\end{keyword}

\end{frontmatter}


\section{Introduction}

The crystalline organic scintillator \textit{trans}-stilbene (hereafter referred to as stilbene) has attracted renewed interest for radiation detection applications \cite{Prasad2018,DiFulvio2017,Goldsmith2016} due to the recent development of a solution-based growth method that enables the fabrication of large monocrystals with excellent neutron-gamma pulse shape discrimination (PSD) and high light output \cite{Carman2013,Zaitseva2015,Bourne2016}.  Stilbene, like many crystalline organic scintillators, exhibits an anisotropic scintillation response to heavy charged particles; the response is dependent on the charged particle trajectory with respect to the crystalline axis.  The light output response has been measured by several authors for a limited number of recoil directions in stilbene crystals \cite{Heckmann1961,Tsukada1962,Brooks1974,Albert1982,Hansen2002,Shimizu2003} and across a full hemisphere \cite{Schuster2017}.  The following is a brief summary of the measurements of interest to this paper. 

The first measurement of the anisotropic response of stilbene was reported in \cite{Heckmann1961}.  The authors determined the crystal orientation using a polarization microscope and reported that the response to 6.93 MeV $\alpha$ particles was at a maximum when traveling along the \textit{b}-axis, a minimum along the \textit{c'}-axis, and a value between the two extrema along the \textit{a}-axis.  A follow-up measurement in \cite{Brooks1974} reported the ratio of the maximum and minimum response for proton recoils at  8 and 22 MeV.  The authors did not report any knowledge of the orientation of the crystalline axes or how the directions of the maximum and minimum response were determined.  The results reported in \cite{Schuster2017} measured proton recoils at 2.5 MeV and 14.1 MeV over a full hemisphere for four stilbene crystals, where two of the crystals had labeled crystalline axes and two did not.  The measurements were reported to be in agreement with \cite{Heckmann1961} with the maximum response for proton recoils along the crystalline \textit{b}-axis and the minimum response for recoils along the \textit{c'}-axis.  The results from \cite{Brooks1974} and \cite{Schuster2017} also show that the magnitude of the response anisotropy decreases as proton recoil energy increases.

While performing measurements of the scintillation response anisotropy of stilbene crystals over a full hemisphere at 11 distinct recoil proton energies between 560 keV and 10 MeV, we found a disagreement between our measurements and reports in the literature regarding the direction of the maximum light output.  This paper presents a subset of our measurements of the scintillation response anisotropy for proton recoils in stilbene, focusing on the light output for recoils along the \textit{a}, \textit{b}, and \textit{c'} axes directions.  Five stilbene crystals were measured, four grown by Inrad Optics and one grown at Lawrence Livermore National Laboratory (LLNL).  The LLNL crystal was previously measured in \cite{Schuster2017} where it was denoted as sample ``Stilbene Cubic A".  The measurements in this paper include the following: (1) measurements of proton recoils between 1.3 and 10 MeV along the  \textit{a}, \textit{b}, and \textit{c'} crystalline axes using scatter kinematics with an 11.3 MeV neutron beam; (2) measurements of proton recoils along the \textit{a} and \textit{b} axes at 2.0 and 2.8 MeV using scatter kinematics with a 4.8 MeV neutron beam; (3) measurements of full energy deposition events along the \textit{a} and \textit{b} axes using a 4.8 MeV neutron beam; and (4) measurements of the count rate for $^{252}$Cf neutrons traveling in the direction of the \textit{a} and \textit{b} axes.  Measurements (1) - (3) were performed at the Triangle Universities Nuclear Laboratory (TUNL) using the tandem Van de Graaff accelerator, and measurement (4) was performed at North Carolina State University.  All of the measurements indicated the \textit{a}-axis as the direction of the maximum light output for proton recoils --- in disagreement with previous literature.

\section{Stilbene structure and axes identification}


The only measurement in the literature of the stilbene light output to also report the methods used for the crystalline axes identification is \cite{Heckmann1961}.  The proper determination and labeling of the crystalline axes was necessary for the measurements presented in this work.  The following describes the methods used by Inrad Optics for stilbene growth and axes identification.  It is assumed that the LLNL crystal's axes were identified using similar methods.

The stilbene crystals were grown from solution \cite{Zaitseva2011a} as single crystals with no visible inclusions or defects.  Stilbene crystallizes in the monoclinic system, with a structure first described in \cite{Robertson1937} and refined in \cite{Finder1974,Bernstein1975,Hoekstra1975}.  The space group and lattice constants in \cite{Bernstein1975} were used as they are consistent with earlier reports of anisotropic behavior in organic scintillators \cite{Heckmann1961}: space group P21/a, \textit{a} = 12.382 \AA, \textit{b} = 5.720 \AA, \textit{c'} = 15.936 \AA, and  $\beta = 114.15^{\circ}$.  Note that \textit{a} in the representation reported in \cite{Finder1974} is not equivalent to the others due to the use of a different space group.

\begin{figure}[!t]
	\centering
	\includegraphics[width= 0.5\linewidth]{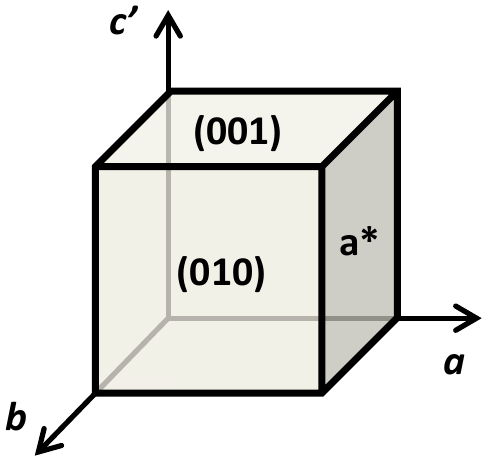}
	\caption{Orientation of the stilbene cubes from Inrad Optics.}
	\label{inrad_stil}
\end{figure}

Solution-grown stilbene boules exhibit two sets of broad natural faces which were identified as ($\bar{2}$03) and (001) by x-ray diffraction; these planes can be quickly differentiated visually due to the much larger birefringence through ($\bar{2}$03) relative to (001).  (010) is orthogonal to the broad natural faces and was confirmed by x-ray diffraction.  The stilbene cubes had faces cut parallel to (010) and (001).  The third set of cube faces is denoted as a*  and the direction normal to (001) is identified as \textit{c'} (refer to Figure \ref{inrad_stil} for an illustration of the face and axis labeling).  

Additional defining characteristics were used to confirm the orientation of the stilbene cubes.  Stilbene cleaves along the (001) \textit{a-b} plane, normal to \textit{c'}.  Stilbene is optically biaxial, with three indices of refraction and two optic axes which together define an optic plane; the cube faces can be quickly verified by examination between crossed polarizers. Extinctions for (010) faces occur at a small angle from the cube edges, since the optic plane is parallel to \textit{b} and rotated away from (001) by approximately $5^{\circ}$.  Looking down \textit{b} reveals strong colors on a* faces.  The final set of faces (001) yield extinctions symmetric with the cube edges.  

\section{Experiment setup at Triangle Universities Nuclear Laboratory}

\begin{figure}[!t]
\begin{subfigure}{0.5\textwidth}
	\centering
	\includegraphics[width= \linewidth]{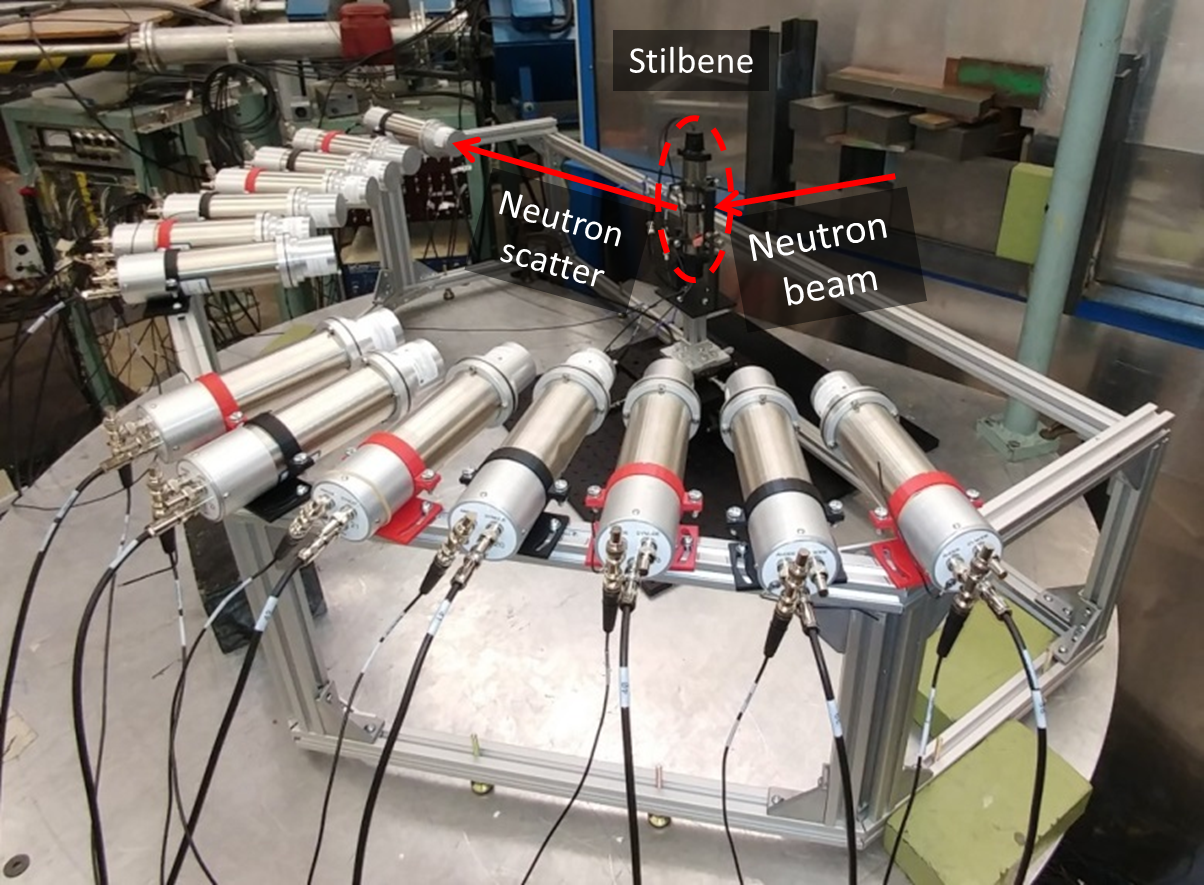}
	\caption{}
	\label{exp_setup_pic}
\end{subfigure}
\hfill
\begin{subfigure}{0.5\textwidth}
 	\centering
	\includegraphics[width= \linewidth]{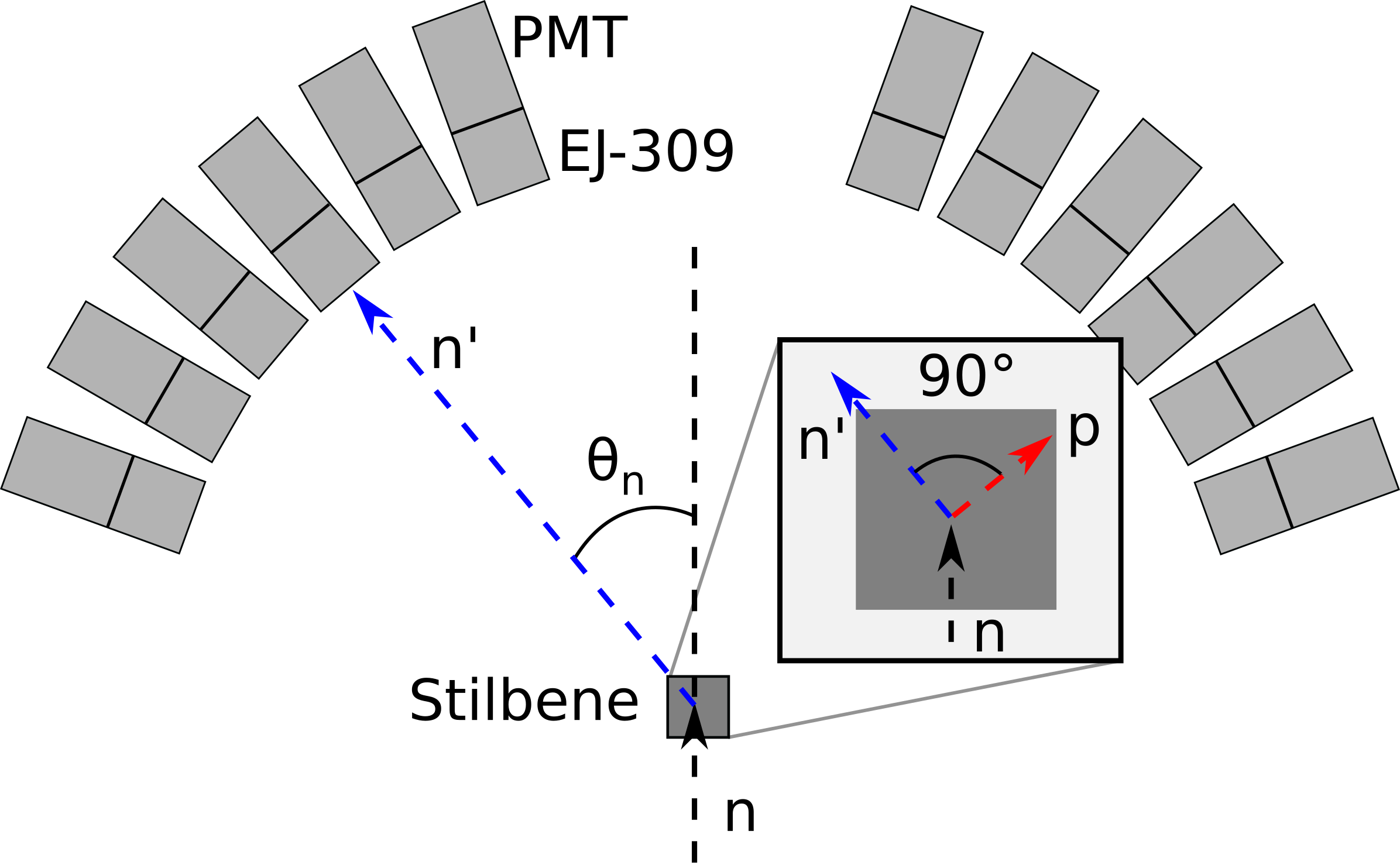}
	\caption{}
	\label{exp_setup_sketch}
\end{subfigure}
\caption{(a) Coincident scatter measurement setup at TUNL and (b) a sketch of the setup.  Neutrons elastically scatter off protons in the stilbene and continue to the EJ-309 liquid scintillator backing detectors.  The directions and energies of proton recoils in the stilbene are calculated using the incident neutron energy and the neutron scatter angle ($\theta_n$) between the stilbene and a backing detector.}
\label{exp_setup}
\end{figure}

The measurements at TUNL were performed using quasi-monoenergetic neutron beams produced with the tandem Van de Graaff accelerator.  Deuterons were accelerated toward a deuterium gas cell to produce neutrons via the D($d$,$n$)$^3$He reaction.  The experiment setup is shown in Figure \ref{exp_setup}.  The measurement relies on the coincident detection of a neutron event in the stilbene detector and one of the 5.08 cm diameter x 5.08 cm long EJ-309 backing detectors, which were aligned at selected angles with respect to the stilbene detector.  The proton recoil energy in the stilbene detector was calculated using:
\begin{equation}
	E_p = E_n \sin^2\theta_n
\end{equation}
where $E_p$ is the proton recoil energy, $E_n$ is the incident neutron energy, and $\theta_n$ is the scatter angle of the neutron.  The direction of the proton recoil in the stilbene crystal  was also determined kinematically, because $\theta_p+\theta_n=90^{\circ}$, where $\theta_p$ is the scatter angle of the proton.  The uncertainty in $\theta_p$ for a single n-p scatter event was dominated by the size of the 5.08 x 5.08 cm EJ-309 backing detectors which were positioned at a nominal distance of 60 cm from the stilbene detector; it was $<4^{\circ}$.  The uncertainty on $E_p$ for a single n-p scatter event was dominated by the energy spread of the quasi-monoenergetic neutron beam and the finite size of the EJ-309 backing detectors; it was $500$ keV.


\subsection{Detectors and data acquisition}

\begin{figure}[!t]
\begin{subfigure}{0.5\textwidth}
  \centering
  \includegraphics[width=0.5\linewidth]{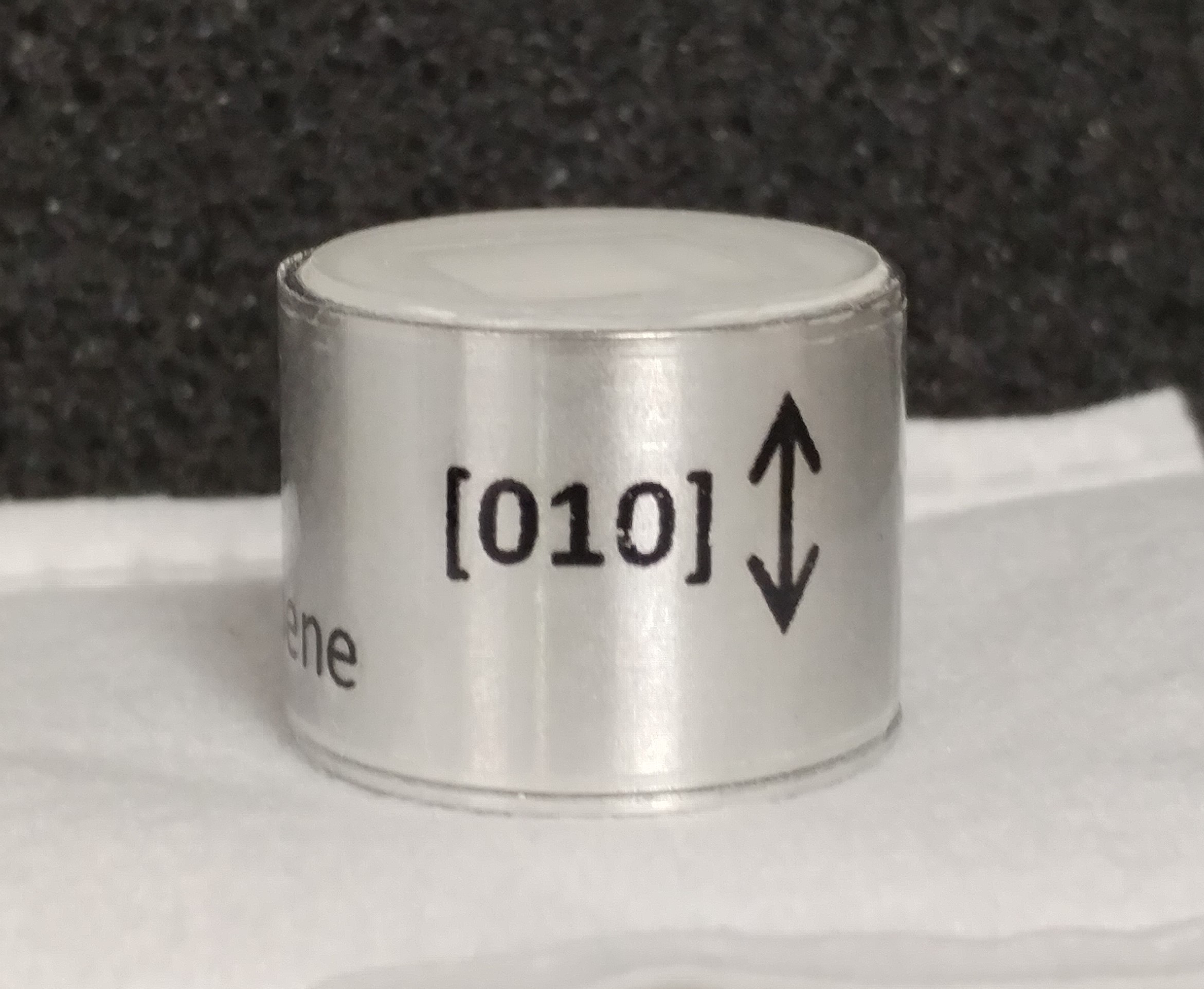}
  \caption{}
  \label{crystal_a}
\end{subfigure}
\hfill
\begin{subfigure}{0.5\textwidth}
  \centering
  \includegraphics[width=0.5\linewidth]{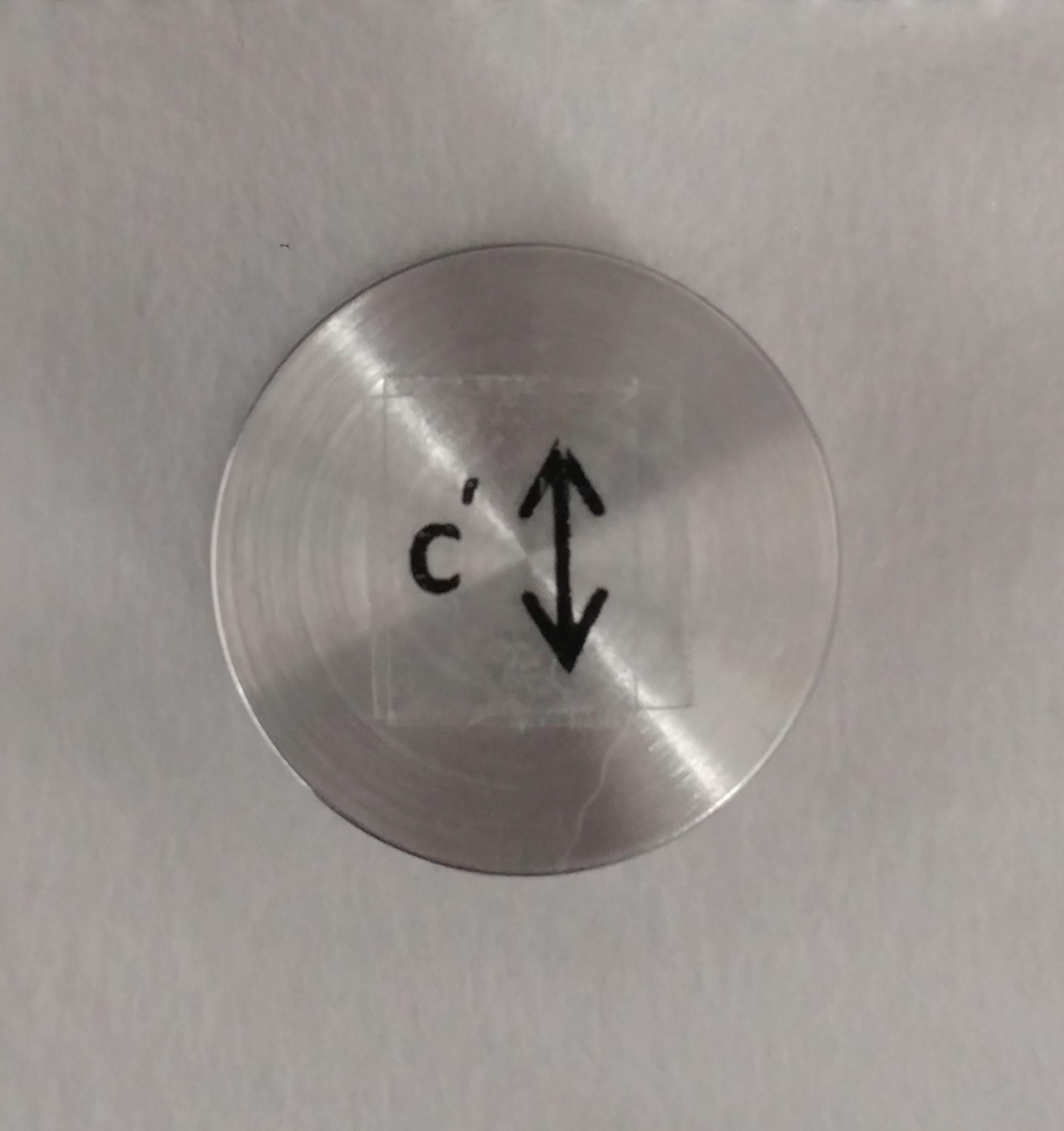}
  \caption{}
  \label{crystal_b}
\end{subfigure}
\caption{Stilbene crystal in aluminum housing with marked axes (note that [010] is equivalent to the \textit{b}-axis direction).  The 1 cc cubic stilbene samples were each enclosed in a cylindrical aluminum housing with a fused silica window.  The region between the stilbene crystal and the aluminum housing was filled TiO$_2$ powder which served as a diffuse reflector, and the crystal was coupled to the silica window with optical-grade silicone grease.   }
\label{stil_crystals}
\end{figure}

Details of the five stilbene crystals are given in Table \ref{stil_details}.  The crystals grown by Inrad Optics were provided with the axes marked as shown in Figure \ref{stil_crystals}.  The LLNL crystal was provided with the \textit{a} and \textit{b} axes marked.  Crystals 1-3 were mounted to three Hamamatsu R7111 28 mm photomultiplier tubes (PMTs), and crystals 4 and 5 were mounted to Electron Tube 9134SB 29 mm PMTs.  Each crystal was mounted to a PMT such that proton recoils in the \textit{a-b}, \textit{a-c'}, or \textit{b-c'} plane could be measured. 

\begin{table}[h]
	\caption{Stilbene crystal details.  ``Recoil plane" refers to the crystalline plane in which proton recoil trajectories were measured for a specific crystal.}
	\label{stil_details}
	\begin{center}
		\begin{tabular}{cccc}
			\hline
			Make & Volume (cc) & Recoil plane & Number \\
			\hline
			\multirow{4}{*}{Inrad} & \multirow{4}{*}{1.00} & \textit{a}-\textit{c'} & 1 \\ 
									             && \textit{b}-\textit{c'} & 2 \\
									             && \textit{a}-\textit{b} & 3\\
										  && \textit{a}-\textit{b} & 4 \\
			\hline
		           LLNL & 6.86 & \textit{a}-\textit{b} & 5\\
			\hline
		\end{tabular}
	\end{center}
\end{table}

\begin{figure}[!t]
	\centering
	\includegraphics[width= 0.5\linewidth]{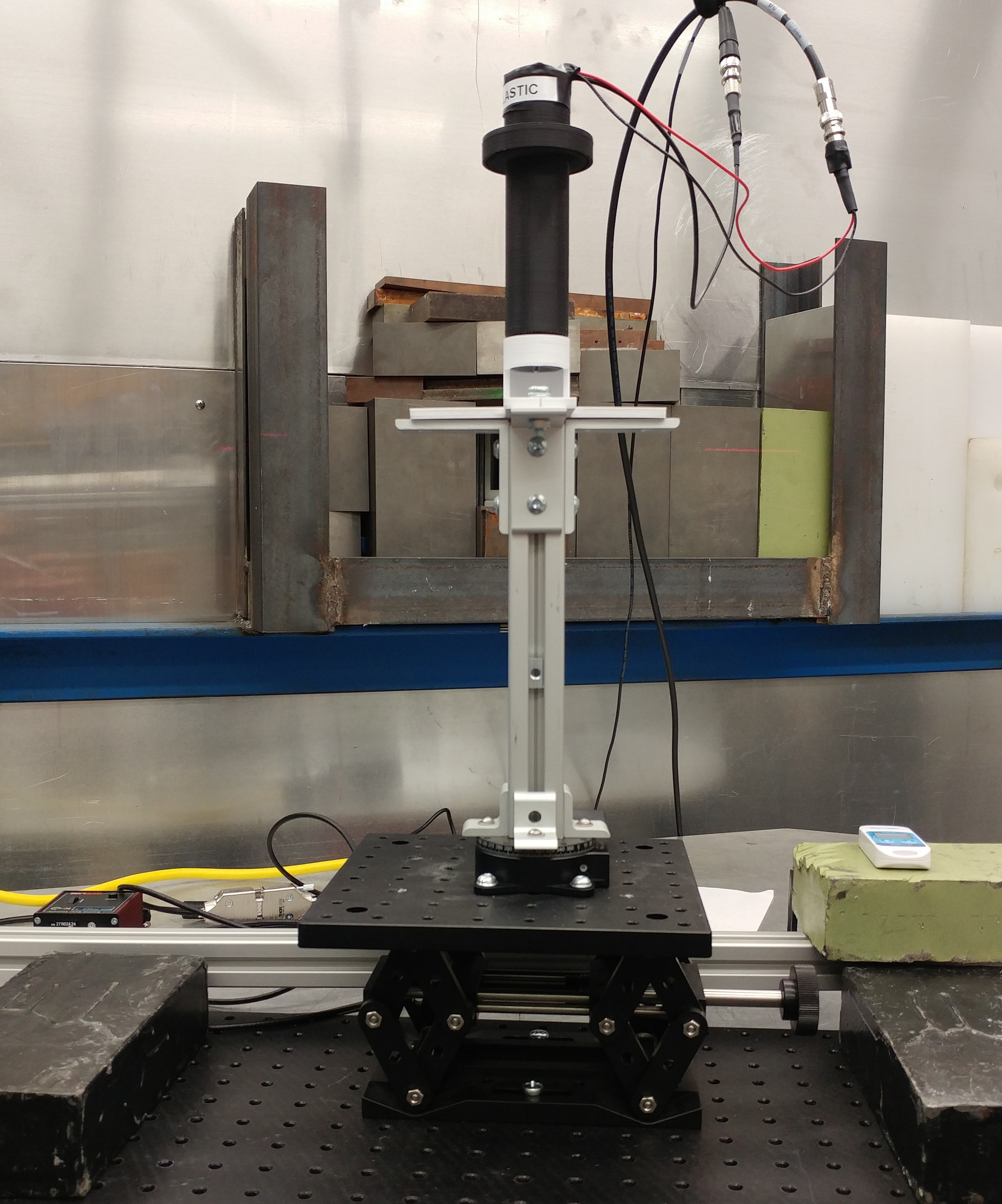}
	\caption{Detector mounted to rotational stage.}
	\label{rot_stage}
\end{figure}

The stilbene detectors were enclosed in light-tight housings, and the PMTs were covered with a cylindrical piece of mu-metal.  The assembled stilbene detectors were mounted to a $360^{\circ}$ rotational table as shown in Figure \ref{rot_stage}.  The crystals were rotated to measure proton recoils along specific trajectories with respect to the crystalline lattice using the backing detectors set at fixed angles.  The 12 backing detectors were mounted to an aluminum frame, positioned with 6 detectors left of the neutron beam with respect to the flight path of the neutrons (i.e. beam left)  and 6 detectors to the right of the beam (i.e. beam right).  The detectors were placed at $20^{\circ}$, $30^{\circ}$, $40^{\circ}$, $50^{\circ}$, $60^{\circ}$, and $70^{\circ}$ relative to the neutron beam direction.  Calibrations were performed using $^{137}$Cs at the beginning and end of the measurements for each stilbene detector.  Detector pulses were recorded using a Struck SIS3316 digitizer.  

\subsection{Plastic scintillator measurements}
Directional measurements of a plastic scintillator were performed to confirm that the response anisotropy is due to effects within the crystalline stilbene detector and not effects from the measurement system. Plastic scintillators do not exhibit directional dependence, as demonstrated in \cite{Schuster2017}, and thus, the absence of light output anisotropy in a plastic sample provides confirmation that the measurement system is not the source of the anisotropy.  A 1 cc EJ-228 plastic scintillator mounted to an Electron Tube 9134SB PMT was characterized using the coincident scatter measurement apparatus.  A similar method as in \cite{Schuster2017} was used, where the observed standard deviation ($\sigma_{obs}$) for measurements at a given energy was compared to the average statistical uncertainty ($\sigma_{stat}$) of those measurements.  The measurements were performed with an 11.3 MeV neutron beam, and the plastic was rotated through 170$^{\circ}$ in 10$^{\circ}$ increments resulting in the measurement of 18 proton recoil directions for 6 proton recoil energies between 1.3 and 10 MeV.  The relative uncertainties ($\sigma_{obs}/\mu$ and $\sigma_{stat}/\mu$) were calculated for each proton recoil energy.  The 1.3 MeV proton recoils showed the poorest agreement between $\sigma_{obs}/\mu$ and $\sigma_{stat}/\mu$ with $\sigma_{obs}/\mu =  0.0127$ and $\sigma_{stat}/\mu = 0.0095$.  The difference between these values is 0.0032 or 0.32\%.  This is significantly less than the 5\% to 30\% difference expected for the stilbene light output, and shows that the light output anisotropy characterized in the stilbene measurements is not caused by an external effect of the measurement system, but rather is due entirely to the light emission within the stilbene crystal.

\section{Quasi-monoenergetic neutron beam measurements} \label{4}

\subsection{11.3 MeV neutron beam} \label{4.1}
\begin{figure}[!t]
	\centering
	\begin{subfigure}{0.45\textwidth}
	  \centering
	  \includegraphics[width=\linewidth]{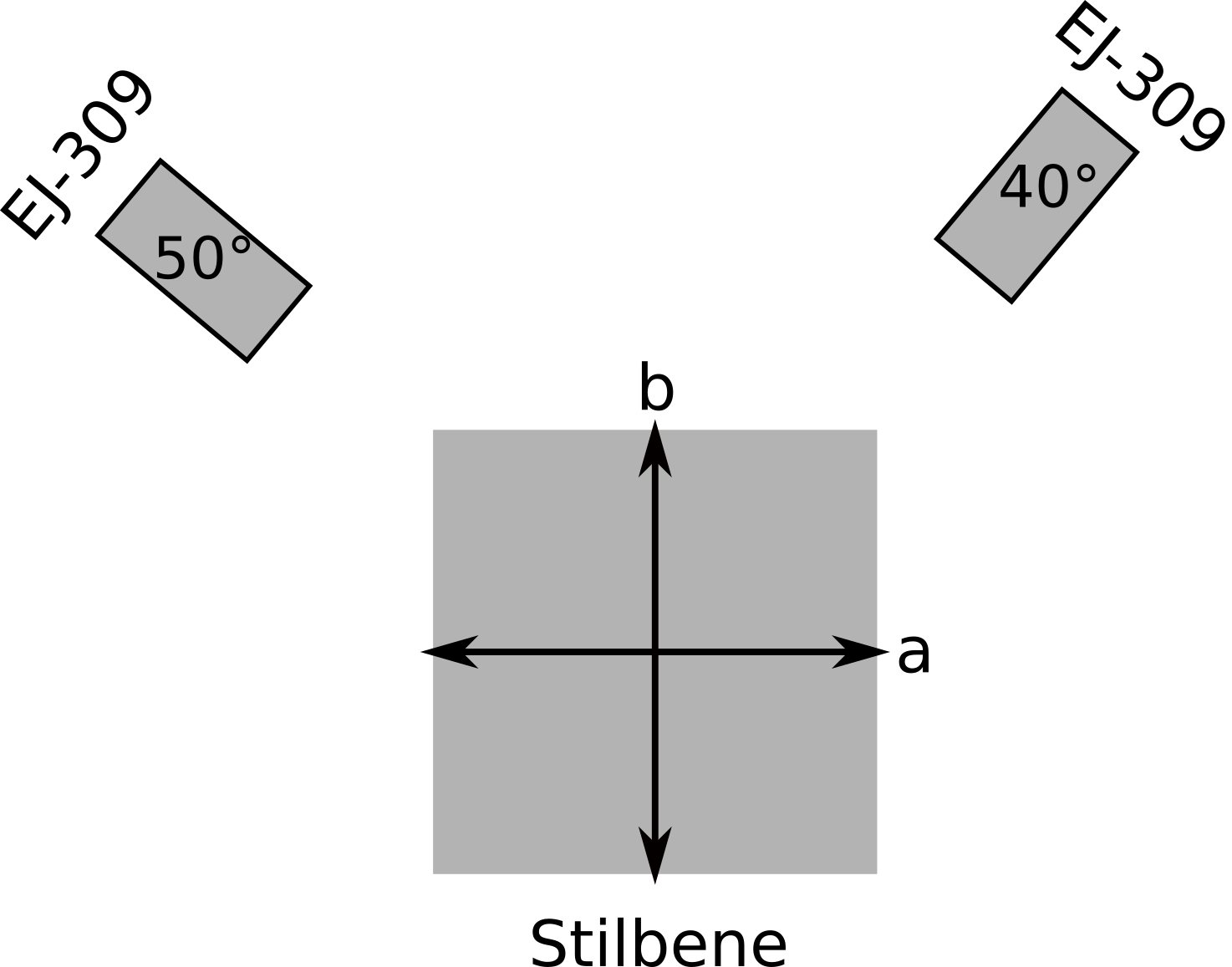}
	  \caption{$\theta_{rot}=0^{\circ}$ }
	  \label{0deg}
	\end{subfigure}
	\hfill
	\begin{subfigure}{0.45\textwidth}
	  \centering
	  \includegraphics[width=\linewidth]{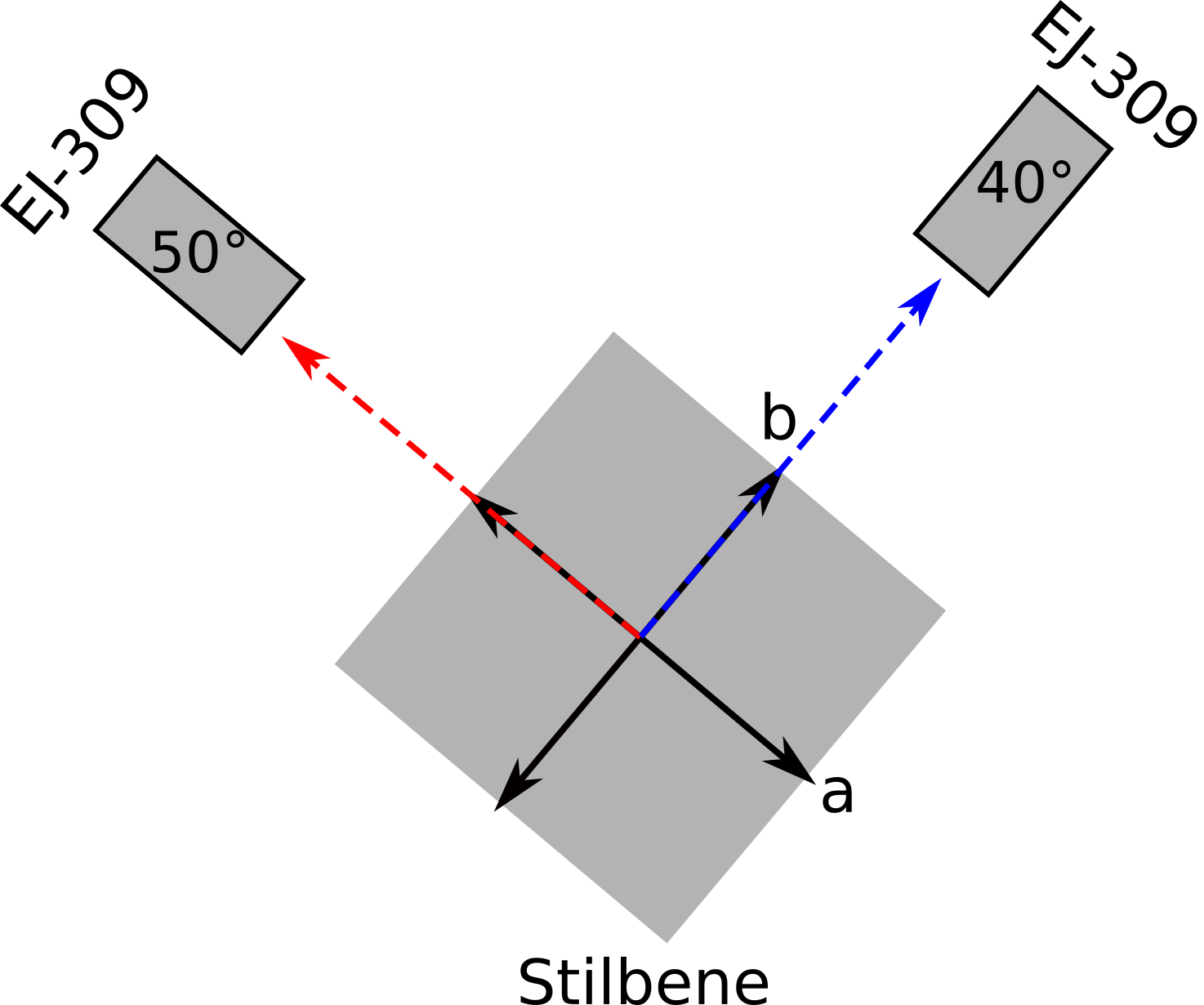}
	  \caption{$\theta_{rot}=40^{\circ}$ }
	  \label{lo}
	\end{subfigure}
	\caption{ Overhead view of the stilbene rotation for the \textit{c'}-vertical crystal orientation (proton recoils in the \textit{a-b} plane) illustrating the positions of the $50^{\circ}$ beam left and $40^{\circ}$ beam right EJ-309 backing detectors (not to scale).  $\theta_{rot}$ is the angle of the rotation stage relative to the direction of the neutron beam.  The marking on the \textit{b}-axis was aligned to the neutron beam and was defined as $\theta_{rot}=0^{\circ}$, shown in (a).   (b) shows the crystal when rotated clock-wise $40^{\circ}$ with the \textit{b}-axis pointing at the $40^{\circ}$ beam right backing detector and the \textit{a}-axis pointing at the $50^{\circ}$ beam left detector.  Following the relationship $\theta_p+\theta_n=90^{\circ}$, neutron scatters into the $40^{\circ}$ detector correspond to proton recoils along the \textit{a}-axis while neutron scatters into the $50^{\circ}$ detector correspond to proton recoils along the \textit{b}-axis.}
	\label{recoil_dir}
\end{figure}

An 11.3 MeV neutron beam was used to individually characterize crystals 1-3 for recoil protons traveling in a fixed plane  (\textit{a-c'} plane, \textit{b-c'} plane, and \textit{a-b} plane) with the setup in Figure \ref{exp_setup}.  Each crystal was mounted to the rotational stage and oriented with either the \textit{a}, \textit{b}, or \textit{c'}-axis in line with the  beam left $70^{\circ}$ backing detector.  The crystals were rotated from the beam left $70^{\circ}$ backing detector to the beam right $70^{\circ}$ backing detector in $10^{\circ}$ increments so that neutron scatter events measured in complimentary backing detectors corresponded to proton recoils along either the \textit{a}, \textit{b}, or \textit{c'}-axis.  Proton recoils along two of the major axes were measured at each rotation angle using beam left and beam right backing detectors as shown in Figure \ref{recoil_dir}.  

\begin{figure}[!t]
  \centering
  \includegraphics[width=0.7\linewidth]{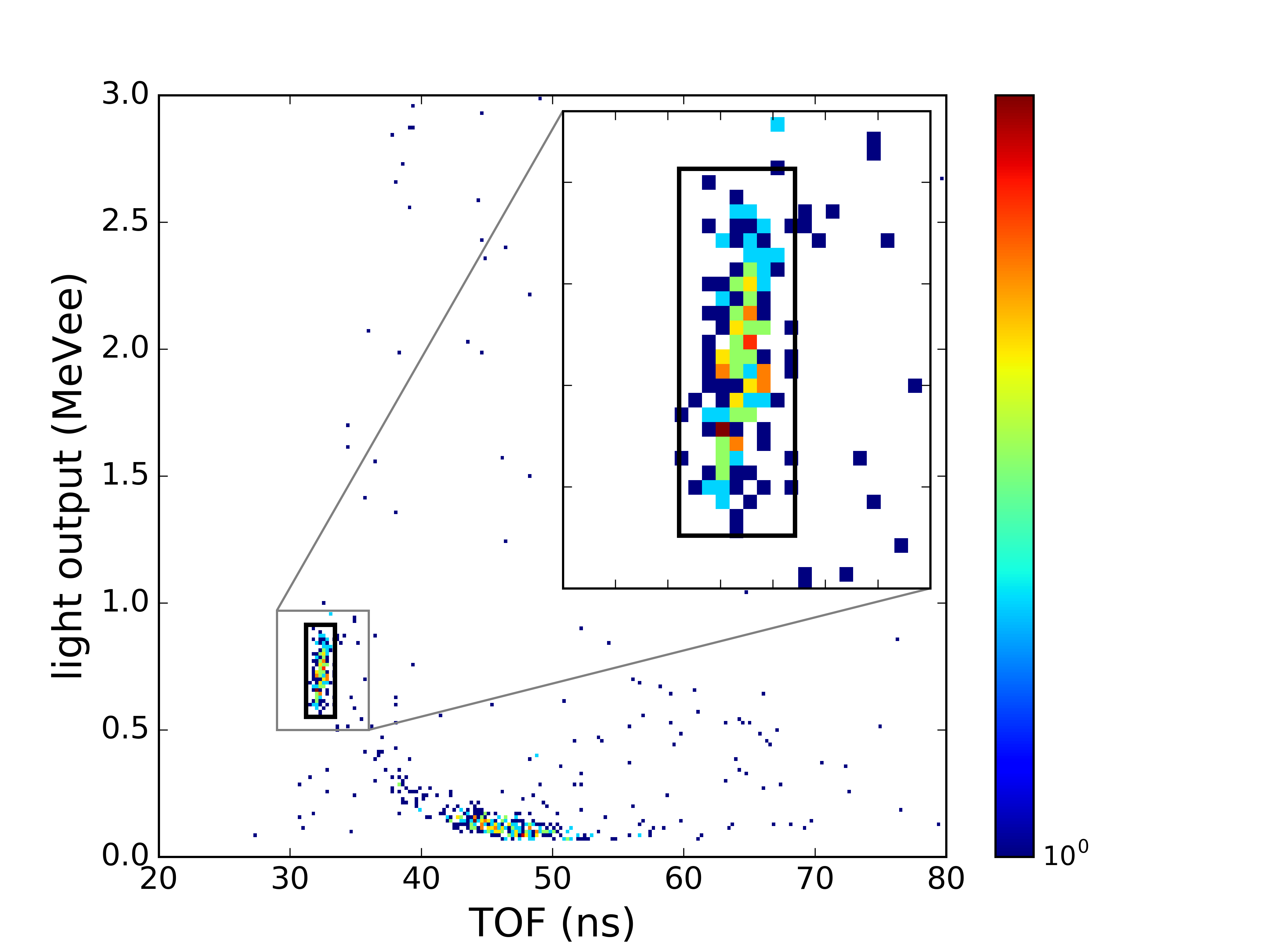}
  \caption{Light output plotted against time-of-flight (TOF) for events recorded in the stilbene detector in coincidence with the $30^{\circ}$ EJ-309 backing detector.  The cluster in the black box corresponds to single n-p scatter events with an energy of 2.8 MeV.  The cluster at lower light output is due to breakup neutrons from the D(d,n+p) reaction.}
  \label{scatter_events}
\end{figure}

The mean light output for proton recoils was calculated by localizing events from single n-p scatters, as shown in Figure \ref{scatter_events}.  The n-p scatter events were identified by examining the light output versus the time-of-flight for a neutron that scattered in the stilbene and then into a backing detector.  The selected events were then projected onto the light output axis so that the mean light output and its uncertainty could be determined.  The mean light output was measured for proton recoils at 1.3, 2.8, 4.7, 6.6, 8.5, and 10.0 MeV (corresponding to neutron scatters into the $20^{\circ}$, $30^{\circ}$, $40^{\circ}$, $50^{\circ}$, $60^{\circ}$, and $70^{\circ}$ backing detectors, respectively)  in each of the 3 crystals for proton recoil directions along the \textit{a}, \textit{b}, or \textit{c'}-axis.

\begin{figure}[!t]
	\centering
	\includegraphics[width= \linewidth]{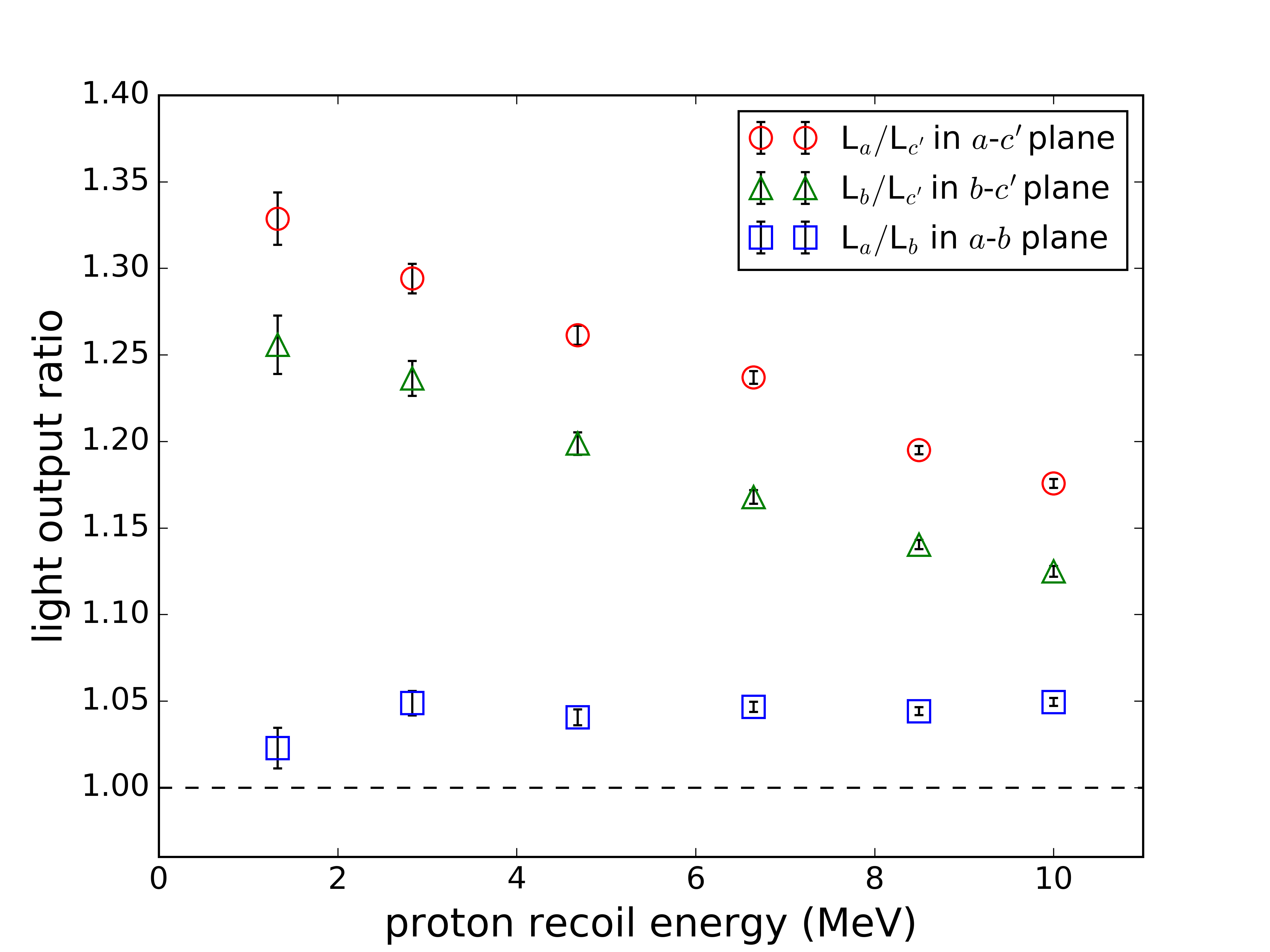}
	\caption{Ratios of the maximum light output recorded in a plane to the minimum light output recorded in that plane.  L$_i$ is the light output for recoil protons traveling along the $i$-axis.   }
	\label{ratio_plot}
\end{figure}

The ratios of the maximum to minimum light output in each plane were calculated and are shown in  Figure \ref{ratio_plot}.  The maximum light output was produced by proton recoils traveling along the \textit{a}-axis.  Note that the ratios in the \textit{a-c'} plane are greater than the ratios in the \textit{b-c'} plane, and the ratios measured in the \textit{a-b} plane are all greater than 1.  The measurements for all three crystals indicate that the maximum light output is for recoils along the \textit{a}-axis.  The expected trend from the literature of decreasing anisotropy with increasing energy is also seen in the ratios for the \textit{a-c'} plane and the \textit{b-c'} plane, which have a maximum at 1.3 MeV and a minimum at 10 MeV, while the ratio for the \textit{a-b} plane is nearly constant.     

\subsection{4.8 MeV neutron beam}
Two additional measurements were made at TUNL with the neutron beam energy at 4.8 MeV using crystals 4 and 5 from Table \ref{stil_details} (hereafter referred to as the Inrad crystal and the LLNL crystal, respectively) oriented for measurements in the \textit{a-b} plane.  The first measurement was performed using the same method as the coincident scatter measurement described in section \ref{4.1}.  The second measurement was performed by measuring full energy deposition events along the \textit{a} and \textit{b} axes.

\subsubsection{Coincident scatter measurement}

The coincident scatter measurement setup shown in Figure \ref{exp_setup} was used to measure proton recoils in the Inrad and LLNL crystals resulting from neutron scatters into the $40^{\circ}$ and $50^{\circ}$ backing detectors with 4.8 MeV incident neutrons.  The $40^{\circ}$ and $50^{\circ}$ backing detectors were chosen due to the higher count rate for scatters in those directions.  Neutron scatters into the $40^{\circ}$ backing detector correspond to proton recoils at 2.0 MeV; neutron scatters into the $50^{\circ}$ backing detectors correspond to proton recoils at 2.8 MeV.  The crystals were rotated such that proton recoils were directed down the \textit{a} and \textit{b} axes for neutron scatters into the beam left and beam right $40^{\circ}$ and $50^{\circ}$ backing detectors.  The results are shown in Table \ref{scatter_table}, and they agree with the conclusion that proton recoils along the \textit{a}-axis produce more light than those along the \textit{b}-axis.  It should also be noted that the measured light output is statistically identical between the Inrad and LLNL crystals for the same recoil energy and recoil axis.

\begin{table}[h]
           \caption{Proton recoil light output for neutron scatters at $40^{\circ}$ (2.0 MeV) and $50^{\circ}$ (2.8 MeV) with statistical uncertainties.}
           \label{scatter_table}
	\begin{center}
		\begin{tabular}{c|c|ccc}
		     \hline
		     Recoil Energy & Crystal & Recoil Axis & L (MeVee) &  L$_a/$L$_b$ \\
		     \hline
		     \multirow{4}{*}{2.0 MeV} & \multirow{2}{*}{Inrad} & \textit{a} & $0.551\pm0.010$  & \multirow{2}{*}{$1.056\pm0.034$}\\
		     && \textit{b} & $0.522\pm0.014$ & \\
		     \cline{2-5}
		     &\multirow{2}{*}{LLNL} & \textit{a} &  $0.560\pm0.009$ & \multirow{2}{*}{$1.057\pm0.023$} \\
		     && \textit{b} & $0.530\pm0.008$\\ 
		     \hline
		     \multirow{4}{*}{2.8 MeV} & \multirow{2}{*}{Inrad} & \textit{a} & $0.930\pm0.011$ & \multirow{2}{*}{$1.059\pm0.017$}\\
		     && \textit{b} & $0.878\pm0.009$ \\
		     \cline{2-5}
		     &\multirow{2}{*}{LLNL} & \textit{a} &  $0.923\pm0.008$ & \multirow{2}{*}{$1.048\pm0.013$}\\
		     && \textit{b} & $0.881\pm0.008$\\ 
		     \hline
		\end{tabular}
	\end{center}
\end{table}

\subsubsection{Full energy deposition measurements}

Full energy deposition events along the \textit{a} and \textit{b} axes of the Inrad and LLNL crystals were measured at TUNL using a 4.8 MeV neutron beam.  The measurements were performed by placing the crystals with the \textit{a} and then the \textit{b} axes in the direction of the neutron beam.  Gamma events were removed in post-processing using a standard charge-integration PSD technique.  All proton recoil events in the stilbene detectors were recorded, and recoils along the forward direction  (full energy deposition scatter events) were identified by fitting the pulse integral spectrum.  Using the same method described in \cite{Schuster2017}, the full energy deposition edge was determined by fitting the measured spectrum to a sloped distribution with a hard cutoff convoluted with a Gaussian resolution function, resulting in the fit function:

\begin{equation}
	\label{ql_edge_fit}
	f(L) = \frac{mL + b}{2} \left[1 - erf\left(\frac{L-\hat{L}}{\sigma\sqrt{2}}\right)\right] - \frac{m\sigma}{\sqrt{2\pi}} e^{\frac{-\left(L-\hat{L}\right)^2}{2\sigma^2}}
\end{equation}

\noindent 
where $m$ and $b$ are the slope parameters, $\hat{L}$ is the edge location, and $\sigma$ is the light output resolution.  

\begin{figure}[!t]
	\centering
	\includegraphics[width= \linewidth]{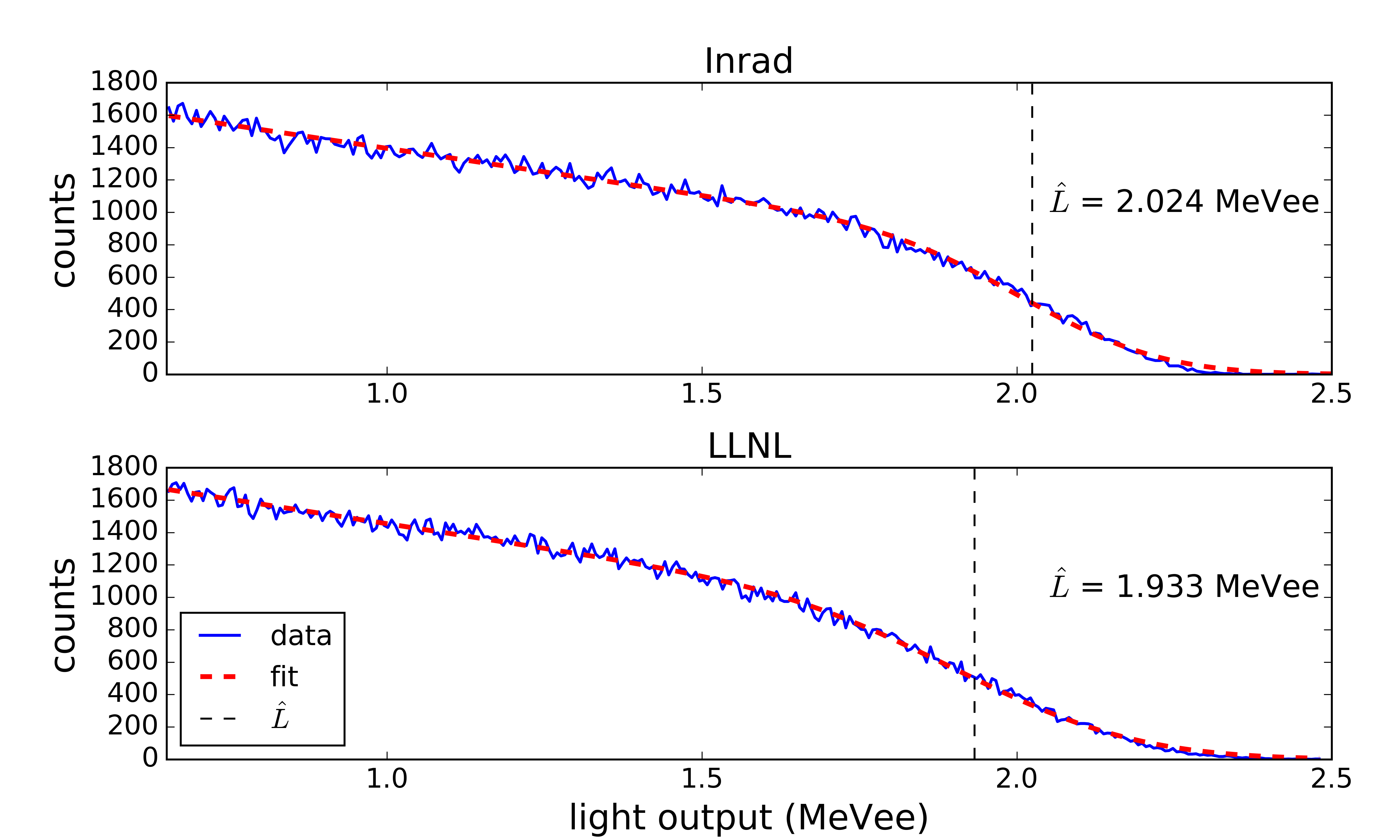}
	\caption{4.8 MeV neutron spectra with fits for proton recoils along the \textit{a}-axis in the Inrad and LLNL crystals.  The black dashed line is the calculated edge location ($\hat{L}$). }
	\label{ql_edge}
\end{figure}

The calculated edge positions are given in Table \ref{full_e_dep_table}.  Again, the results for both crystals show that the proton recoils along the \textit{a}-axis produce more light than along the \textit{b}-axis.  Examples of measured spectra and the fits are shown in Figure \ref{ql_edge}.  The difference in the calculated edge positions between the Inrad and LLNL crystal is due to multiple scatter events in the much larger LLNL crystal.  Neutrons in larger volume scintillators have a higher probability for multiple scatters.  Due to the nonlinear, concave-up relationship between light output and energy, it is impossible for a multiple scatter event to produce as much light as a full energy deposition single scatter.  This results in a larger number of lower light output events near the full energy deposition edge and a corresponding decrease in the fitted edge position.  This effect only influences the calculation of the full energy deposition edge; it does not occur in the coincident scatter measurements.  Multiple scatter events are mostly discriminated out in the coincident scatter measurements due to the differences in total light output and the time-of-flight for multiple scatter events compared to single scatter events.    

\begin{table}[h]
           \caption{4.8 MeV full energy deposition calculated edge positions with the statistical uncertainties of the fits.}
           \label{full_e_dep_table}
	\begin{center}
		\begin{tabular}{c|ccc}
		     \hline
		     Crystal & Axis & $\hat{L}$ (MeVee) &  $\hat{L}_a/\hat{L}_b$ \\
		     \hline
		     \multirow{2}{*}{Inrad} & \textit{a} & $2.024\pm0.009$ & \multirow{2}{*}{$1.038\pm0.006$} \\
		     & \textit{b} & $1.950\pm0.007$ \\
		     \hline
		     \multirow{2}{*}{LLNL} & \textit{a} &  $1.933\pm0.008$ & \multirow{2}{*}{$1.038\pm0.006$}\\
		     & \textit{b} & $1.862\pm0.007$\\ 
		     \hline
		\end{tabular}
	\end{center}
\end{table}

\section{$^{252}$Cf measurements}

\begin{figure}[!t]
	\centering
	\includegraphics[width= 0.7\linewidth]{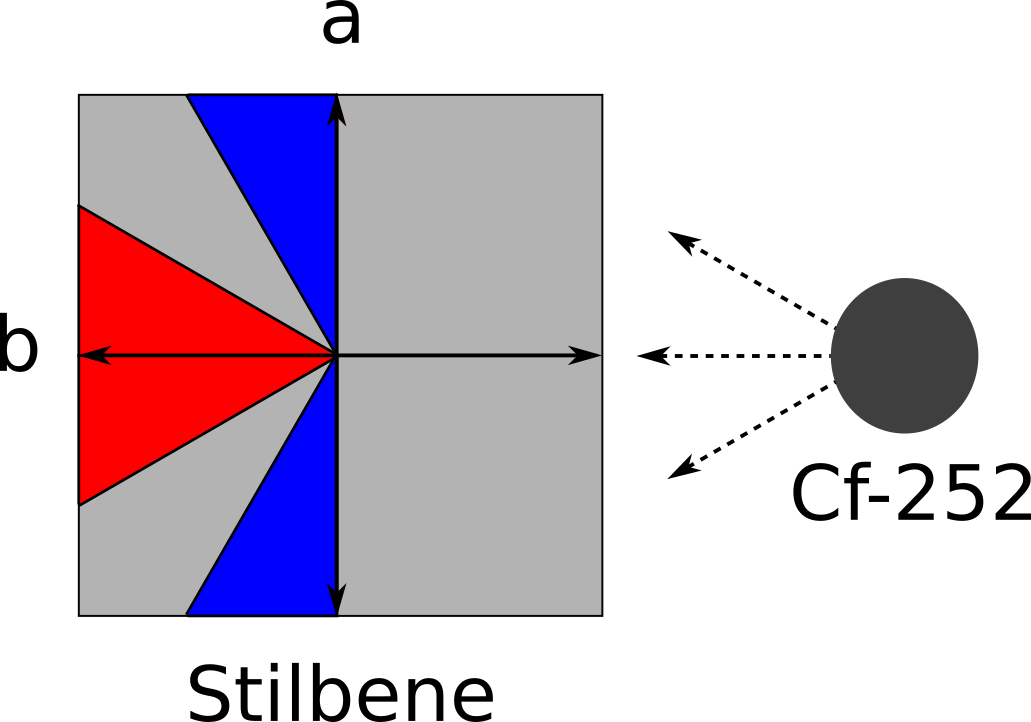}
	\caption{Sketch of the $^{252}$Cf measurement setup.  The blue shaded areas represent possible neutron scatter directions and the red shaded area represents possible proton recoil directions.  The threshold, energy of the incident neutron, and direction of the incident neutron determine possible scatter directions (area of shaded regions). The shaded areas of this sketch are for the specific case of a neutron with energy slightly greater than threshold (approximately 3 MeV) and traveling along the \textit{b}-axis. }
	\label{cf_setup}
\end{figure}

A $^{252}$Cf neutron source provides another, simple way to measure the light output anisotropy in the stilbene crystals.  The recoil direction resulting in the maximum light output will also have the maximum recorded count rate if the threshold and distance from the source are fixed.  Proton recoils with energies near the threshold, traveling near the direction of the maximum light output will produce enough scintillation photons to be recorded as an event, while recoils along other directions will not, resulting in a higher count rate along the direction of the maximum light output. It should be noted that the protons in these measurements do not recoil strictly along the \textit{a} and \textit{b} axes directions as in the other measurements in this work.  The proton recoil directions were partially directed by setting the threshold of the detectors to 1.0 MeVee, which corresponds to a recoil proton energy of approximately 3 MeV.  A sketch of this is shown in Figure \ref{cf_setup}.  The 1.0 MeVee threshold restricted the detected proton recoils to a cone about the axis that is colinear with the source-detector axis or ``in-line" with the source (e.g the \textit{b}-axis is in-line with the source in Figure \ref{cf_setup}).   The angle of the cone is dependent on the incident neutron energy and direction.   

The measurements were performed by placing a $^{252}$Cf source at a fixed distance from the Inrad and LLNL stilbene crystals with the \textit{b}-axis in-line with the source such that protons would recoil in a direction within a cone about \textit{b}-axis.  The count rate was recorded.  The crystals were then rotated 90$^{\circ}$ such that the \textit{a}-axis was in-line with the source and proton recoils were directed within a cone about the \textit{a}-axis, and the count rate was again recorded.  A standard charge-integration PSD technique was applied in post-processing to separate neutron and gamma events.  The gamma count was used to normalize the neutron count for comparison between measurements, since gamma interactions in crystalline organic scintillators do not exhibit scintillation anisotropy \cite{Schuster2016a}.    

The ratios of the count rates when the \textit{a}-axis was in-line with the source (recoils in the direction of the \textit{a}-axis) to when the \textit{b}-axis was in-line with the source (recoils in the direction of the \textit{b}-axis) are shown in Table \ref{cf_table}.  The count rate ratios were greater than 1 for both crystals.  These results confirm that recoils along the \textit{a}-axis correspond to the direction of the maximum light output, in agreement with section \ref{4} and in disagreement with previous results in the literature.

\begin{table}[h]
           \caption{$^{252}$Cf count rate ratios.  $N_a/N_b$ is the ratio of the number of events recorded when the \textit{a}-axis was in-line with the source to the number of events recorded when the \textit{b}-axis was in-line with the source.}
           \label{cf_table}
	\begin{center}
		\begin{tabular}{c|c}
		     \hline
		     Crystal & $N_a/N_b$ \\
		     \hline
		      Inrad & $1.043\pm0.014$\\
		      LLNL & $1.051\pm0.008$\\
		     \hline
		\end{tabular}
	\end{center}
\end{table}

\section{Comparison of the magnitude of change in light output}
With the \textit{a}-axis established as the direction of the maximum light output for proton recoil events, the light output ratios for recoils in the \textit{a-c'} plane (circles in Figure \ref{ratio_plot}) can be compared with the magnitude of change in the light output measured in \cite{Brooks1974} and \cite{Schuster2017}.  The results in \cite{Schuster2017} define the magnitude of change of the light output as the ratio of the maximum to the minimum light output for a crystal: 

\begin{equation}
	\label{A_L}
	A_L = \frac{\hat{L}_{max}}{\hat{L}_{min}}
\end{equation}

\begin{figure}[!t]
	\centering
	\includegraphics[width= \linewidth]{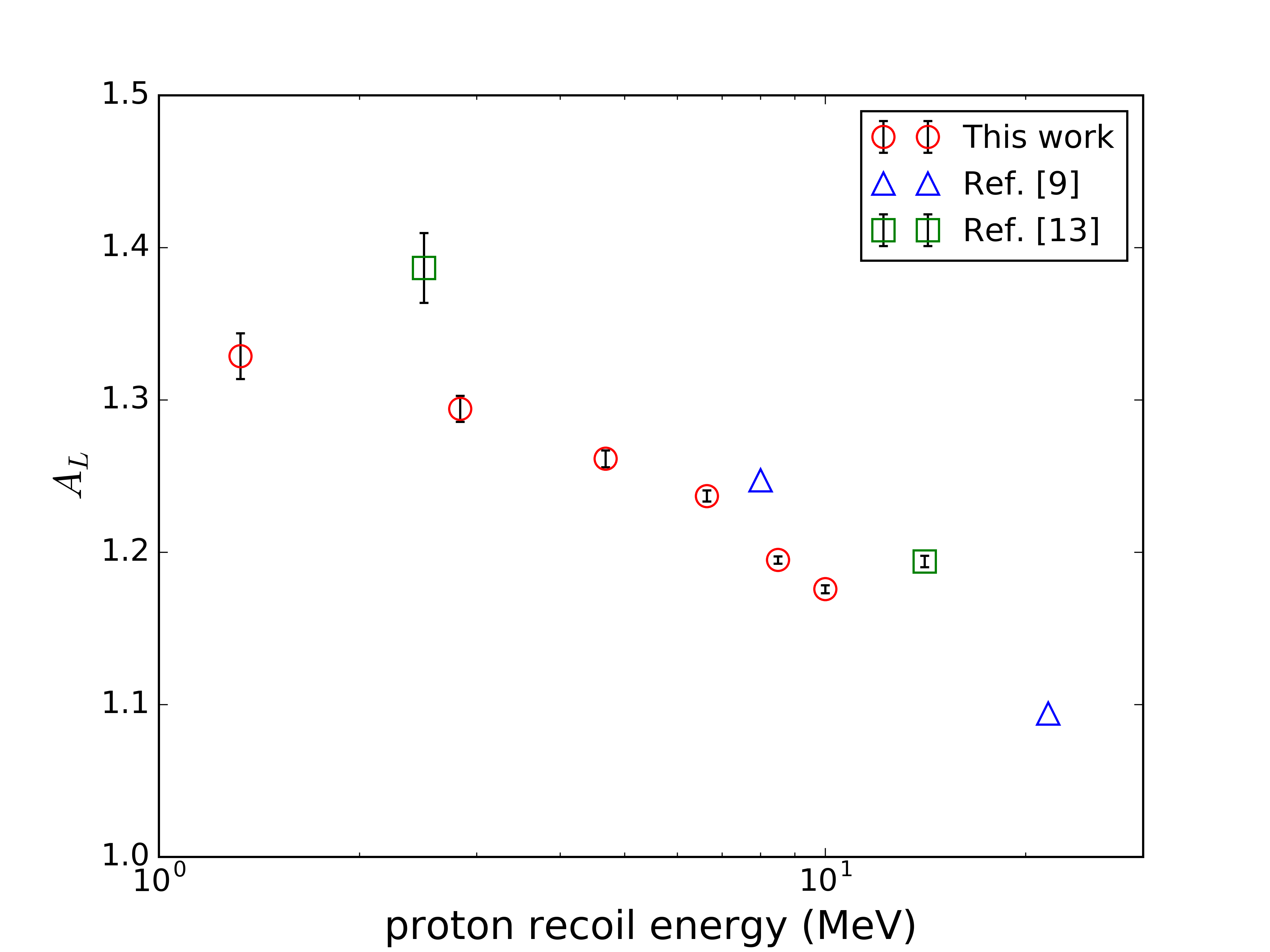}
	\caption{Magnitude of change in the light output response vs energy as reported in \cite{Brooks1974}, \cite{Schuster2017}, and this work.  The data points from 				    \cite{Schuster2017} were calculated by averaging the reported ratios for three solution grown stilbene crystals. }
	\label{compare}
\end{figure} 

\noindent
Figure \ref{compare} shows the light output ratio for recoils along the \textit{a}-axis and \textit{c'}-axis plotted with the ratios of maximum to minimum light output measured in \cite{Brooks1974} and \cite{Schuster2017}.  The ratios from this work are lower than those previously reported.  One potential cause of this effect is that these measurements were performed using the R7111 PMTs which were partially saturating for proton recoils above 3 MeV.  The saturation becomes more pronounced as the proton recoil energy increases resulting in a decreased ratio because the effect is larger for recoils along the direction of the maximum response than other recoil directions. Also note that the coincident scatter technique used in this work is more accurate than the edge fitting technique, particularly at lower energies where the full energy deposition edge has poorer resolution.  As noted in \cite{Schuster2017}, changes in the binning or the fitted light output range result in significant changes in the fitted edge position.   
  
\section{Discussion of disagreement with previous literature}
The conclusion that the \textit{a}-axis is the direction of the maximum light output is in disagreement with several authors \cite{Heckmann1961,Brooks1974,Schuster2017}, as discussed previously.  Subsequent to the completion of the measurements presented in this paper, the results reported in \cite{Schuster2017} were reviewed by the author who is a co-author of this paper.  It was determined that an error was made in the labeling of the \textit{a} and \textit{b} axes that resulted in the axes being swapped.  It has been concluded that the \textit{a}-axis is the direction of maximum light output measured in \cite{Schuster2017}, and that the measurements are in agreement with the results of this paper.

The measurements reported in \cite{Brooks1974}, while of some use for comparison, should not be considered precise due to the omission of uncertainties on the measurements and the absence of an explanation of the method used to determine the maximum and minimum directions of the light output.  In addition, the crystalline axis orientation unknown, and the maximum and minimum light output directions reported in \cite{Heckmann1961} were assumed correct without independent validation.  Based on the information provided in \cite{Brooks1974}, it is possible that the $A_L$ values reported for stilbene were the ratios of the response for recoils along the \textit{a} and \textit{c'} axes.  

The measurements reported in \cite{Heckmann1961} were the first to show the \textit{b}-axis as the direction of the maximum light output,  but were performed with $\alpha$ particles unlike the other measurements discussed in this paper.  The crystalline axis orientation was determined using a polarization microscope, making it unlikely that the axes were incorrectly identified.  Differences in the direction of the maximum anisotropy for protons and $\alpha$ particles may be a real effect, and further research into the effect of the stopping power on the anisotropic response could be of interest.  It is also possible that the measurements were influenced by the polycrystalline nature of early melt-grown stilbene samples.  A polycrystalline stilbene sample was measured in \cite{Tsukada1962}, and the direction of the maximum light output was reported to nearly coincide with the \textit{c}-axis.  This displays the large affect poorly grown samples had on scintillation measurements.

\section{Conclusion}
Measurements of the light output anisotropy for proton recoils in stilbene were conducted for recoils traveling along the \textit{a}, \textit{b}, and \textit{c'} axes, and we have concluded, contrary to previous literature, that the direction of the maximum light output for proton recoils in stilbene is along the \textit{a}-axis.  The \textit{c'}-axis was confirmed as the direction of the minimum light output, in agreement with previous publications.  The measurements were in agreement across five different stilbene crystals including four grown by Inrad Optics and one grown by LLNL.  The $^{252}$Cf measurement provides a simple method for measuring the directional response anisotropy without the need for an accelerator or neutron generator and can be used to quickly determine the crystalline orientation.  This work contains preliminary results of the scintillation response characterization of stilbene crystals over a full hemisphere.  Future work will include full hemisphere characterizations with over 100 proton recoil directions at 11 distinct energies leading to a full characterization of the stilbene scintillation response as a function of the proton recoil energy and scatter direction between 560 keV and 10 MeV.

\section*{Acknowledgments}
The authors would like to thank the TUNL staff that helped to support these measurements, and in particular John Dunham for his help with operating the tandem.

This work was sponsored in part by the NNSA Office of Defense Nuclear Nonproliferation R\&D through the Consortium for Verification Technology (CVT) grant number DE-NA0002534 and the Nuclear Regulatory Commission through the North Carolina State University Graduate Fellowship in Nuclear Engineering. 

\section*{References}

\bibliography{directional_anisotropy_in_stilbene_v6}

\begin{thebibliography}{10}
\expandafter\ifx\csname url\endcsname\relax
  \def\url#1{\texttt{#1}}\fi
\expandafter\ifx\csname urlprefix\endcsname\relax\def\urlprefix{URL }\fi
\expandafter\ifx\csname href\endcsname\relax
  \def\href#1#2{#2} \def\path#1{#1}\fi

\bibitem{Prasad2018}
M.~K. Prasad, {Safeguards Technology Development Program 1st Quarter FY 2018
  Report}\href {http://dx.doi.org/10.2172/1417284}
  {\path{doi:10.2172/1417284}}.

\bibitem{DiFulvio2017}
A.~{Di Fulvio}, T.~H. Shin, T.~Jordan, C.~Sosa, M.~L. Ruch, S.~D. Clarke, D.~L.
  Chichester, S.~A. Pozzi,
  \href{http://dx.doi.org/10.1016/j.nima.2017.02.082}{{Passive assay of
  plutonium metal plates using a fast-neutron multiplicity counter}}, Nuclear
  Instruments and Methods in Physics Research, Section A: Accelerators,
  Spectrometers, Detectors and Associated Equipment 855~(February) (2017)
  92--101.
\newblock \href {http://dx.doi.org/10.1016/j.nima.2017.02.082}
  {\path{doi:10.1016/j.nima.2017.02.082}}.
\newline\urlprefix\url{http://dx.doi.org/10.1016/j.nima.2017.02.082}

\bibitem{Goldsmith2016}
J.~E.~M. Goldsmith, M.~D. Gerling, J.~S. Brennan, D.~J. Throckmorton, J.~I.
  Helm, System construction of the stilbene compact neutron scatter camera\href
  {http://dx.doi.org/10.2172/1331427} {\path{doi:10.2172/1331427}}.

\bibitem{Carman2013}
L.~Carman, N.~Zaitseva, H.~P. Martinez, B.~Rupert, I.~Pawelczak, A.~Glenn,
  H.~Mulcahy, R.~Leif, K.~Lewis, S.~Payne,
  \href{http://dx.doi.org/10.1016/j.jcrysgro.2013.01.019}{{The effect of
  material purity on the optical and scintillation properties of solution-grown
  trans-stilbene crystals}}, Journal of Crystal Growth 368 (2013) 56--61.
\newblock \href {http://dx.doi.org/10.1016/j.jcrysgro.2013.01.019}
  {\path{doi:10.1016/j.jcrysgro.2013.01.019}}.
\newline\urlprefix\url{http://dx.doi.org/10.1016/j.jcrysgro.2013.01.019}

\bibitem{Zaitseva2015}
N.~Zaitseva, A.~Glenn, L.~Carman, H.~{Paul Martinez}, R.~Hatarik, H.~Klapper,
  S.~Payne, \href{http://dx.doi.org/10.1016/j.nima.2015.03.090}{{Scintillation
  properties of solution-grown trans-stilbene single crystals}}, Nuclear
  Instruments and Methods in Physics Research, Section A: Accelerators,
  Spectrometers, Detectors and Associated Equipment 789 (2015) 8--15.
\newblock \href {http://dx.doi.org/10.1016/j.nima.2015.03.090}
  {\path{doi:10.1016/j.nima.2015.03.090}}.
\newline\urlprefix\url{http://dx.doi.org/10.1016/j.nima.2015.03.090}

\bibitem{Bourne2016}
M.~M. Bourne, S.~D. Clarke, N.~Adamowicz, S.~A. Pozzi, N.~Zaitseva, L.~Carman,
  \href{http://dx.doi.org/10.1016/j.nima.2015.10.025}{{Neutron detection in a
  high-gamma field using solution-grown stilbene}}, Nuclear Instruments and
  Methods in Physics Research, Section A: Accelerators, Spectrometers,
  Detectors and Associated Equipment 806 (2016) 348--355.
\newblock \href {http://dx.doi.org/10.1016/j.nima.2015.10.025}
  {\path{doi:10.1016/j.nima.2015.10.025}}.
\newline\urlprefix\url{http://dx.doi.org/10.1016/j.nima.2015.10.025}

\bibitem{Heckmann1961}
P.~H. Heckmann, H.~Hansen, A.~Flammersfeld,
  \href{http://link.springer.com/10.1007/BF01342470}{{Die
  Richtungsabhaengigkeit der Szintillations-Lichtausbeute von duennen
  Anthrazen- und Stilben-Kristallen beim Beschuss mit alpha-Strahlen}},
  Zeitschrift fuer Physik 162~(1) (1961) 84--92.
\newblock \href {http://dx.doi.org/10.1007/BF01342470}
  {\path{doi:10.1007/BF01342470}}.
\newline\urlprefix\url{http://link.springer.com/10.1007/BF01342470}

\bibitem{Tsukada1962}
K.~Tsukada, S.~Kikuchi,
  \href{http://www.sciencedirect.com/science/article/pii/0029554X62900071}{{Directional
  anisotropy in the characteristics of the organic-crystal scintillators}},
  Nuclear Instruments and Methods 17~(3) (1962) 286--288.
\newblock \href {http://dx.doi.org/10.1016/0029-554X(62)90007-1}
  {\path{doi:10.1016/0029-554X(62)90007-1}}.
\newline\urlprefix\url{http://www.sciencedirect.com/science/article/pii/0029554X62900071}

\bibitem{Brooks1974}
F.~D. Brooks, D.~T.~L. Jones, {Directional anisotropy in organic scintillation
  crystals}, Nuclear Instruments and Methods 121~(1) (1974) 69--76.
\newblock \href {http://dx.doi.org/10.1016/0029-554X(74)90141-4}
  {\path{doi:10.1016/0029-554X(74)90141-4}}.

\bibitem{Albert1982}
D.~Albert, U.~Bruckner, W.~Hansen, W.~Vogel, {Determination of light efficiency
  of stilbene scintillators and their application to in-core spectrometry of
  fast neutrons}, Nuclear Instruments and Methods in Physics Research Section
  A: Accelerators, Spectrometers, Detectors and Associated Equipment 200 (1982)
  397--402.

\bibitem{Hansen2002}
W.~Hansen, D.~Richter, {Determination of light output function and angle
  dependent correction for a stilbene crystal scintillation neutron
  spectrometer}, Nuclear Instruments and Methods in Physics Research, Section
  A: Accelerators, Spectrometers, Detectors and Associated Equipment 476~(1-2)
  (2002) 195--199.
\newblock \href {http://dx.doi.org/10.1016/S0168-9002(01)01430-9}
  {\path{doi:10.1016/S0168-9002(01)01430-9}}.

\bibitem{Shimizu2003}
Y.~Shimizu, M.~Minowa, H.~Sekiya, Y.~Inoue, {Directional scintillation detector
  for the detection of the wind of WIMPs}, Nuclear Instruments and Methods in
  Physics Research, Section A: Accelerators, Spectrometers, Detectors and
  Associated Equipment 496~(2-3) (2003) 347--352.
\newblock \href {http://dx.doi.org/10.1016/S0168-9002(02)01661-3}
  {\path{doi:10.1016/S0168-9002(02)01661-3}}.

\bibitem{Schuster2017}
P.~Schuster, E.~Brubaker,
  \href{http://dx.doi.org/10.1016/j.nima.2016.11.016}{{Characterization of the
  scintillation anisotropy in crystalline stilbene scintillator detectors}},
  Nuclear Instruments and Methods in Physics Research, Section A: Accelerators,
  Spectrometers, Detectors and Associated Equipment 859~(November 2016) (2017)
  95--101.
\newblock \href {http://dx.doi.org/10.1016/j.nima.2016.11.016}
  {\path{doi:10.1016/j.nima.2016.11.016}}.
\newline\urlprefix\url{http://dx.doi.org/10.1016/j.nima.2016.11.016}

\bibitem{Zaitseva2011a}
N.~Zaitseva, L.~Carman, A.~Glenn, J.~Newby, M.~Faust, S.~Hamel, N.~Cherepy,
  S.~Payne,
  \href{http://dx.doi.org/10.1016/j.jcrysgro.2010.10.139}{{Application of
  solution techniques for rapid growth of organic crystals}}, Journal of
  Crystal Growth 314~(1) (2011) 163--170.
\newblock \href {http://dx.doi.org/10.1016/j.jcrysgro.2010.10.139}
  {\path{doi:10.1016/j.jcrysgro.2010.10.139}}.
\newline\urlprefix\url{http://dx.doi.org/10.1016/j.jcrysgro.2010.10.139}

\bibitem{Robertson1937}
J.~M. Robertson, I.~Woodward,
  \href{http://rspa.royalsocietypublishing.org/cgi/doi/10.1098/rspa.1937.0203}{{X-Ray
  Analysis of the Dibenzyl Series. IV. Detailed Structure of Stilbene}},
  Proceedings of the Royal Society A: Mathematical, Physical and Engineering
  Sciences 162~(911) (1937) 568--583.
\newblock \href {http://dx.doi.org/10.1098/rspa.1937.0203}
  {\path{doi:10.1098/rspa.1937.0203}}.
\newline\urlprefix\url{http://rspa.royalsocietypublishing.org/cgi/doi/10.1098/rspa.1937.0203}

\bibitem{Finder1974}
C.~J. Finder, M.~G. Newton, N.~L. Allinger,
  \href{http://scripts.iucr.org/cgi-bin/paper?a10860{\%}5Cnhttp://scripts.iucr.org/cgi-bin/paper?S0567740874002913}{{An
  improved structure of trans -stilbene}}, Acta Crystallographica Section B
  Structural Crystallography and Crystal Chemistry 30~(2) (1974) 411--415.
\newblock \href {http://dx.doi.org/10.1107/S0567740874002913}
  {\path{doi:10.1107/S0567740874002913}}.
\newline\urlprefix\url{http://scripts.iucr.org/cgi-bin/paper?a10860{\%}5Cnhttp://scripts.iucr.org/cgi-bin/paper?S0567740874002913}

\bibitem{Bernstein1975}
J.~Bernstein,
  \href{http://scripts.iucr.org/cgi-bin/paper?a12142{\%}5Cnhttp://scripts.iucr.org/cgi-bin/paper?S0567740875005031}{{Refinement
  of trans-stilbene: a comparison of two crystallographic studies}}, Acta
  Crystallographica Section B Structural Crystallography and Crystal Chemistry
  31~(5) (1975) 1268--1271.
\newblock \href {http://dx.doi.org/10.1107/S0567740875005031}
  {\path{doi:10.1107/S0567740875005031}}.
\newline\urlprefix\url{http://scripts.iucr.org/cgi-bin/paper?a12142{\%}5Cnhttp://scripts.iucr.org/cgi-bin/paper?S0567740875005031}

\bibitem{Hoekstra1975}
A.~Hoekstra, P.~Meertens, A.~Vos,
  \href{http://scripts.iucr.org/cgi-bin/paper?S0567740875008953}{{Refinement of
  the crystal structure of trans-stilbene (TSB). The molecular structure in the
  crystalline and gaseous phases}}, Acta Crystallographica Section B Structural
  Crystallography and Crystal Chemistry 31~(12) (1975) 2813--2817.
\newblock \href {http://dx.doi.org/10.1107/S0567740875008953}
  {\path{doi:10.1107/S0567740875008953}}.
\newline\urlprefix\url{http://scripts.iucr.org/cgi-bin/paper?S0567740875008953}

\bibitem{Schuster2016a}
P.~Schuster, E.~Brubaker, {Investigating the Anisotropic Scintillation Response
  in Anthracene through Neutron, Gamma-Ray, and Muon Measurements}, IEEE
  Transactions on Nuclear Science 63~(3) (2016) 1942--1954.
\newblock \href {http://dx.doi.org/10.1109/TNS.2016.2542589}
  {\path{doi:10.1109/TNS.2016.2542589}}.

\end{thebibliography}

\end{document}